\documentclass[apj]{emulateapj}
\usepackage{apjfonts}
\usepackage[tight]{subfigure}
\usepackage{amsmath}

\shorttitle{Turbulence in a 3D deflagration model for type Ia SNe:
  I. Scaling properties} 
\shortauthors{F. Ciaraldi-Schoolmann et al.}

\begin{document}

\title{Turbulence in a three-dimensional deflagration model for\\ Type
  Ia supernovae: I. Scaling properties} 

\author{F. Ciaraldi-Schoolmann, W. Schmidt, J. C. Niemeyer,}
\affil{Lehrstuhl f\"ur Astronomie und Astrophysik, Universit\"at
  W\"urzburg, Am Hubland, D-97074 W\"urzburg, Germany} 

\author{F. K. R\"opke and W. Hillebrandt}
\affil{Max-Planck-Institut f\"ur Astrophysik,
  Karl-Schwarzschild-Str. 1, D-85741 Garching, Germany} 

\begin{abstract}
We analyze the statistical properties of the turbulent velocity field
in the deflagration model for Type Ia supernovae. In particular, we
consider the question of whether turbulence is isotropic and consistent
with the Kolmogorov theory at small length scales. 
Using numerical data from a high-resolution simulation of a
thermonuclear supernova explosion (R\"opke et al., 2007), spectra of
the turbulence energy and velocity structure functions are computed. 
We show that the turbulent velocity field is isotropic at small length
scales and follows a scaling law that is consistent with the
Kolmogorov theory until most of the nuclear fuel is burned. At length
scales greater than a certain characteristic scale that agrees with
the prediction of Niemeyer and Woosley (1997), turbulence becomes
anisotropic. Here, the radial velocity fluctuations follow the scaling
law of the Rayleigh-Taylor instability, whereas the angular component
still obeys Kolmogorov scaling. In the late phase of the
explosion, this characteristic scale drops below the numerical resolution of the
simulation.
The analysis confirms that a subgrid-scale model for the unresolved
turbulence energy is required for the consistent calculation of the
flame speed in deflagration models of Type Ia supernovae, and that the
assumption of isotropy on these scales is appropriate. 
\end{abstract}


\section{Introduction}
\label{Intro}

The mechanism of thermonuclear explosions of white dwarf (WD) stars,
giving rise to Type Ia supernovae (SNe~Ia) is still not
understood in full detail.
For three reasons, a key to this problem is the understanding of turbulent
thermonuclear combustion in the deflagration phase---the phase of subsonic
flame propagation which commences the
explosion process. First, it is required for the
correct modeling of the flame propagation in the deflagration model of
SNe~Ia \citep[see][for a review of SN~Ia explosion scenarios]{hillebrandt2000a}. Second, in alternative models, burning
starts out in the deflagration mode, and this is an essential
ingredient to the overall explosion process. Third, turbulence in the
deflagration phase sets the conditions for a possible
deflagration-to-detonation transition (DDT) in the delayed detonation
scenario \citep{roepke2007d,woosley2007a}. The necessary insight into
the details of the turbulent combustion process,
however, is hampered by the fact that full-star simulations of
thermonuclear supernova explosions cannot resolve the structure of the
deflagration flame. At the large scales accessible to simulations, the
flame propagation is determined by flame instabilities and
turbulence. These effects significantly boost the effective burning
speed and, in the pure deflagration model of SNe~Ia, lead to the flame
acceleration required to explode the WD
\citep{reinecke2002d,GamKhok03}. In order to describe the flame
propagation in such simulations, the interaction of the flame with
turbulent velocity fluctuations has to be modeled. These modeling
approaches yield an effective flame propagation speed on the
numerically resolved scales---the so-called turbulent burning speed.

The problem of calculating this turbulent flame propagation speed in
three-dimensional simulations of thermonuclear supernova
explosions\footnote{For a review, see \cite{RoepSchm08}.}
has been the subject of a lively debate. 
One school of thought holds the view that the effective propagation
speed in the flamelet regime would naturally be given by the velocity
scale $v_{\mathrm{RT}}(\ell)\propto \left(g_{\mathrm{eff}}\ell\right)^{1/2}$
associated with the Rayleigh-Taylor (RT) instability induced by
buoyancy in the gravitational field $g_{\mathrm{eff}}$ for any length
scale $\ell$ \citep{Sharp84}. As a subgrid scale model ($\ell =
\Delta$, where $\Delta$ is the numerical cutoff length), this scaling
relation is easily implemented and appears to be motivated by the
basic physics of thermonuclear combustion in Type Ia supernovae
\citep{GamKhok03}. In opposition to this view, \citet{NieHille95} and
\citet{NieKerst97} argued that inevitably turbulent velocity
fluctuations $v'(\ell)$ are dominated by the turbulent cascade at
length scales $\ell$ small compared to the scale of energy injection
by the RT instability and, hence, should follow the Kolmogorov scaling
$v'(\ell)\propto\ell^{1/3}$. \citet{NieWoos97} estimated the
transition length $\ell_{\mathrm{K/RT}}$ between the RT-dominated
length scales (the ``large scales'' $\ell\gtrsim \ell_{\mathrm{K/RT}}$)
and the regime of the turbulent cascade (the ``small scales''
$\ell\lesssim \ell_{\mathrm{K/RT}}$) to be of the order
$10\,\mathrm{km}$. Since the cutoff length $\Delta$ in contemporary
numerical simulation is comparable to or less than
$\ell_{\mathrm{K/RT}}$, it follows that a subgrid scale (SGS) model
for the consistent calculation of the turbulent flame speed has to be
based on the turbulence energy associated with the length scale
$\Delta$, which is determined by the the dynamics of the turbulent
cascade. Such an
SGS model was proposed by \citet{NieHille95} and further developed by
\citet{SchmNie06}. It is not clear \emph{a priori} that the turbulence
to be captured by the SGS ansatz is of Kolmogorov-type and therefore the
approach of \citet{SchmNie06} is not based on this assumption.
However, Kolmogorov scaling is an obvious possibility that has to be
considered.

The best way to gain insight into the properties of turbulence in the
deflagration stage is to analyze it directly in high-resolution simulations of
the deflagration model. Here the effects of gravity and spherical
expansion of the background are naturally accounted for.
In this paper, we present such an \emph{a posteriori} analysis of
turbulence based on 
data from a recent numerical simulation \citep{RoepHille07} which was
carried out on a very
large grid ($1024^{3}$ cells). The characteristics of this
deflagration model as well as derived synthetic observables are in
reasonable agreement with the observations of dimmer (but still
normal) observed SNe~Ia \citep{RoepHille07}. Therefore, our
analysis is based on data from a model that is expected to give a realistic
picture of turbulence in SNe~Ia. The failure of the pure deflagration
scenario to reproduce the brighter end of the SN~Ia sample does not
limit the significance of our results as in all alternative scenarios
currently under discussion a similar deflagration phase initiates the
explosion and sets the stage for the later evolution.

In the simulation analyzed here, the co-moving grid technique introduced by
\citet{Roepke05} allowed for a very small initial cutoff length
$\Delta\sim 1\mathrm{km}$ in the inner regions of the exploding white
dwarf. Although $\Delta$ was gradually increased in the course of the
simulation, it was possible to investigate the behaviour of turbulent
velocity fluctuations at length scales
$\ell\sim{\mathrm{10}}\,\mathrm{km}$ by means of computing kinetic
energy spectrum functions of the velocity field in subdomains selected
by an appropriate window function. The results indicated Kolmogorov
scaling \citep{RoepHille07}. In this article, we refine this analysis
by decomposing the velocity field into radial and angular
components. In addition, we subtract the spherically averaged radial
velocity in order to separate the velocity fluctuations from the mean
radial expansion of the white dwarf. For both components of the
fluctuating velocity field, a Fourier analysis is carried out to
compute kinetic energy spectra and to investigate possible
anisotropies. Since Fourier transforms are distorted by the
non-periodic boundary of the computational domain, we calculate
structure functions of the fluctuating velocity field to obtain
reliable estimates of the scaling properties for the whole dynamical
range of the simulation.

The methodology of our analysis is explained in detail in the
following section. As will be shown in Section~\ref{Results}, RT
scaling is found for the radial fluctuating velocity component at
length scales greater than $\ell_{\mathrm{K/RT}}$, whereas Kolmogorov
scaling applies for smaller length scales. In close agreement with the
prediction by \citet{NieWoos97}, the numerically determined value of
$\ell_{\mathrm{K/RT}}$ is about $15\,\mathrm{km}$ after the onset of
the explosion. The angular velocity component, on the other hand,
closely follows the Kolmogorov scaling law at \emph{all} length
scales. Furthermore, the velocity fluctuations are nearly isotropic
at length scales smaller than $\ell_{\mathrm{K/RT}}$. In the last
section, we discuss possible caveats of our analysis and comment on
the implications for numerical simulations of Type Ia supernovae. 

\section{Analysis of the turbulent velocity field}
\label{Analysis}

For the statistical analysis of turbulence, we have to take into
account that the velocity field is a superposition of turbulent
velocity fluctuations and the bulk expansion of the white dwarf.
To estimate the bulk expansion, we average the radial
component of the velocity field over spherical shells of discrete radii $r_i$:
\begin{equation}
\label{Eq:1}
	\bar{v}(r_i) = \frac{1}{N_{i}}\cdot\sum^{N_{i}}_{j=1}\mathbf{v}(\mathbf{r}_ij)\cdot\mathbf{e}_{r}(\mathbf{r}_ij),
\end{equation}
where $r_i^{2} = i\Delta^{2}(t)$ is an integer multiple of the squared
size $\Delta(t)$ of the grid cells at time $t$. The sum is over all cells in the cubic grid that are located
at the distance $r_i$ from the center, and $N_{i}$ is the corresponding number of cells.
The unit vector in radial direction at the position $\mathbf{r}_ij$ of the $j$-th cell in the $i$-th shell is
denoted by $\mathbf{e}_{r}(\mathbf{r}_ij)$. Using this estimate, we subtract the spherically averaged component $\bar{v}(r_i)$ from the original velocity field. In the following, it is understood that the symbol $\mathbf{v}$ refers to the fluctuating part of the velocity field.

Since the RT-Instability evolves in the direction of gravity, we
perform all computations with velocity components parallel and
perpendicular to the gravitational field. These components correspond
to the radial and angular directions, because the gravitational field
is assumed to be spherically symmetric in the simulation. Thus, we
define $\mathbf{v}_{\|}:=v_{r}\mathbf{e}_{r}$, where $\mathbf{e}_{r}=\mathbf{r}/r$
and $v_r=\mathbf{v}\cdot\mathbf{e}_{r}$,  and 
$\mathbf{v}_\bot:=\mathbf{v}-\mathbf{v}_\|$. The corresponding energy spectrum
functions are obtained by integrating the kinetic energy per unit mass
over spheres of radius $k$ in Fourier space: 
\begin{subequations} 
\label{Eq:3}
\begin{eqnarray}
        E_{\|}(k)&=\frac{1}{2}\oint\mathrm{d}\Omega_{k}\,k^{2}|\hat{\mathbf{v}}_\|(\mathbf{k})|^{2}, \\
        E_{\bot}(k)&=\frac{1}{2}\oint\mathrm{d}\Omega_{k}\,k^{2}|\hat{\mathbf{v}}_\bot(\mathbf{k})|^{2},      
\end{eqnarray}
\end{subequations}
where $\hat{\mathbf{v}}_\|(\mathbf{k})$ and $\hat{\mathbf{v}}_\bot(\mathbf{k})$
are the Fourier transforms of the longitudinal and transversal
velocity components, respectively. For developed turbulence, it is
expected that the the energy spectrum functions follow power laws,
$E(k)\propto k^{-\beta}$, in the inertial subrange of wave numbers. A
disadvantage of Fourier transforms is that the contributions from
small wave numbers are distorted by the non-periodic boundaries of the
computational domain. For this reason, we apply Gaussian window
functions to the data sets as described in \cite{RoepHille07}. Since
the data windowing corresponds to a high-pass filter in Fourier space,
the range of the energy spectra is constrained to higher wave
numbers. 

In contrast to the energy spectra, structure functions are two-point
velocity correlation functions computed in position space. While the
Fourier transforms are performed in Cartesian coordinate systems, we
use spherical coordinates for the computation of structure functions,
which is convenient to define directions parallel and perpendicular to
gravity. Moreover, the computation can be constrained to the interiors
of spheres containing a certain amount of burned matter. We define the
radial velocity increment by the difference of $\mathbf{v}_\|$, i.~e.,
the velocity component in the direction of gravity, at two different
positions $\mathbf{r}_{1}$ and $\mathbf{r}_{2}$, 
\begin{subequations}
\begin{eqnarray}
\delta\mathbf{v}_{\mathrm{rad}}=v_{\|}(\mathbf{r}_{2}) - v_{\|}(\mathbf{r}_{1})
\label{Eq:4a}
\end{eqnarray}
The angular velocity increment is defined by the difference of the
velocities projected in the directions perpendicular to gravity, i~e.,
\begin{eqnarray}
\delta\mathbf{v}_{\mathrm{ang}}=v_\bot(\mathbf{r}_{2}) - v_\bot(\mathbf{r}_{1})
\label{Eq:4b}
\end{eqnarray}
\end{subequations}

Note that these increments do not correspond to longitudinal and
transversal velocity increments, because the velocity components are
generally not parallel or perpendicular to the spatial separation
$\mathbf{r}_{2}-\mathbf{r}_{1}$. However, we think that the above
definitions of velocity increments are better suited to the physics of
RT-driven turbulence in thermonuclear supernovae. Our proposition is
corroborated by the the scaling properties of the structure functions
that will be presented in the following Section. 

The radial and angular structure functions of order $p$ are defined by
the averages of the radial and angular velocity increments to the
power $p$, respectively: 

\begin{subequations}
\begin{eqnarray}
\label{Eq:5a}
S_{p,\mathrm{rad}}(\ell)&=\left\langle|\delta\mathbf{v}_{\mathrm{rad}}|^p\right\rangle,\\
\label{Eq:5b}
S_{p,\mathrm{ang}}(\ell)&=\left\langle|\delta\mathbf{v}_{\mathrm{ang}}|^p\right\rangle,
\end{eqnarray}
\end{subequations}
where the length scale $\ell:=|\mathbf{r}_{2}-\mathbf{r}_{1}|$. There is a
large range of length scales which encompasses the turbulent interior
of the exploding WD. For fully developed turbulence the
structure functions are given by power laws $S_{p,\mathrm{rad}}(\ell)\propto\ell^{\zeta_{p,\mathrm{rad}}}$ and $S_{p,\mathrm{ang}}(\ell)\propto\ell^{\zeta_{p,\mathrm{ang}}}$, where $\zeta_{p,\mathrm{rad}}$ and $\zeta_{p,\mathrm{ang}}$ are characteristic scaling exponents.
If we further assume isotropy,
$\zeta_{p,\mathrm{rad}}\simeq\zeta_{p,\mathrm{ang}}$, and 
$\zeta_p=p/3$ according to the theoretical analysis by
\citet{Kolmogorov41}. In particular, it follows that the turbulent velocity fluctuation $v'(\ell)\propto\ell^{1/3}$. For the turbulent flow driven by the RT instability, on the other hand,
$v'(\ell)\propto\ell^{1/2}$ \citep{DavTay50} corresponding to $\zeta_p=p/2$.

For the numerical computation of the structure
functions, one has to take a sufficient large number of sample points
that are distributed with uniform probability within a spherical
region of prescribed radius in order to achieve converged
statistics. This was achieved by a Monte-Carlo-type algorithm, where
the total number of sample points was varied and, thereby, convergence
was established. 

In order to analyze the isotropy of the velocity field close to the
flame, we performed calculations in small boxes intersected by the
flame. The algorithm is based on the analysis performed by
\citet{Zingale05}. To simplify the calculation, the box is placed
along a coordinate axis, in our case the $z$-axis, which defines the
local direction of gravity. For each cell within the box, the velocity
difference $\delta\mathbf{v}$ between the local velocity and the velocity
at the center of the box is calculated. Then projected contours of the
velocity differences in Fourier space can be constructed, by
integrating $\delta\hat{\mathbf{v}}(\mathbf{k})$ over circles of radius
$k_\rho= \sqrt{k^2_x + k^2_y}$ in planes perpendicular to the
$z$-component of $\mathbf{k}$. This procedure was applied for several
positions of the box center corresponding to different fractions of
burned matter in the box. 

\begin{figure*}
\centering
        \subfigure[t = 0.5 seconds]
        {\includegraphics[width=8.00cm]{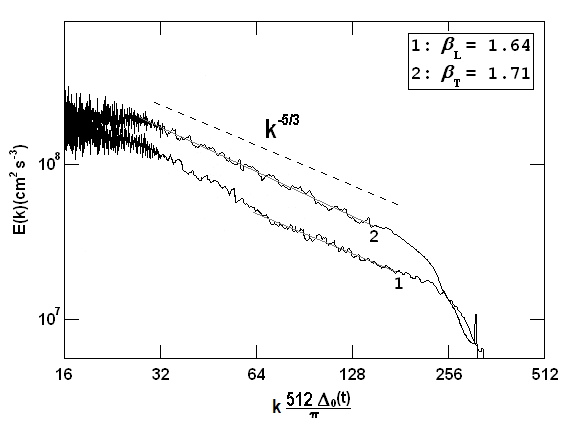}}
        \subfigure[t = 1.0 seconds]
        {\includegraphics[width=8.00cm]{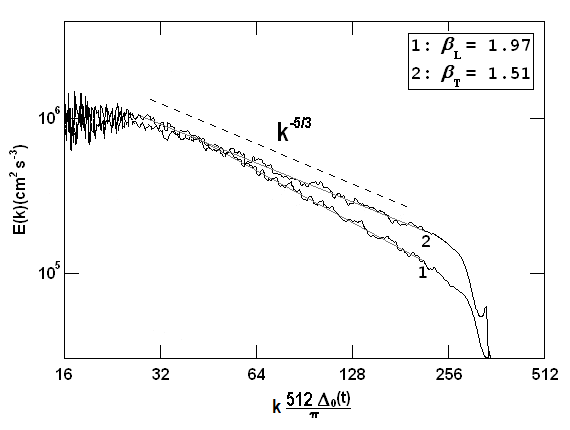}}
\caption{Longitudinal (1) and transversal (2) energy spectrum function (black curve) at t = 0.5 (a) and t = 1.0 (b) seconds in comparison with the Kolmogorov energy spectrum (dashed line).} 
\label{Fig:1}
\end{figure*}
 
\begin{figure*}[t]
\centering
        \subfigure[t = 0.3 seconds]
        {\includegraphics[width=5.20cm]{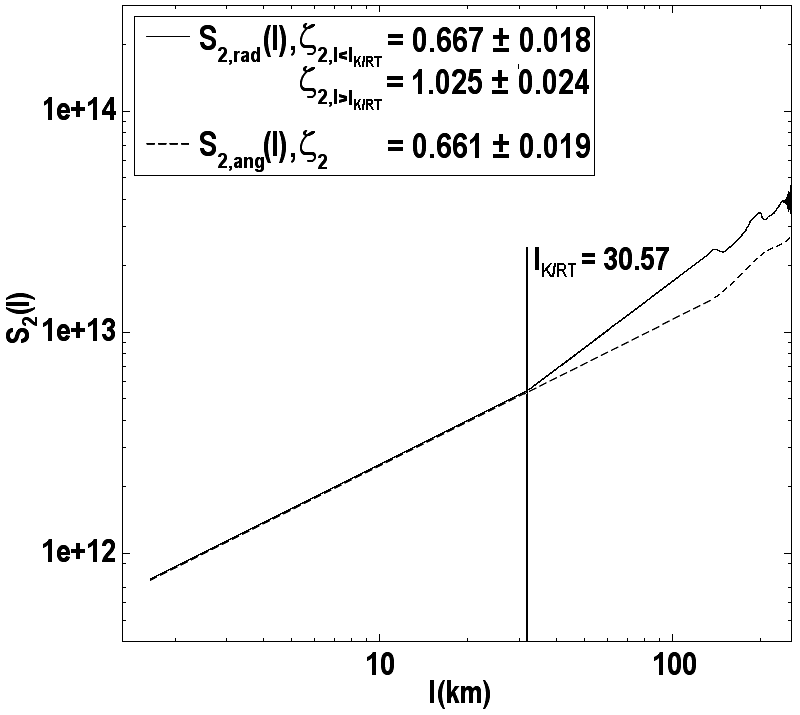}}
        \subfigure[t = 0.4 seconds]
        {\includegraphics[width=5.20cm]{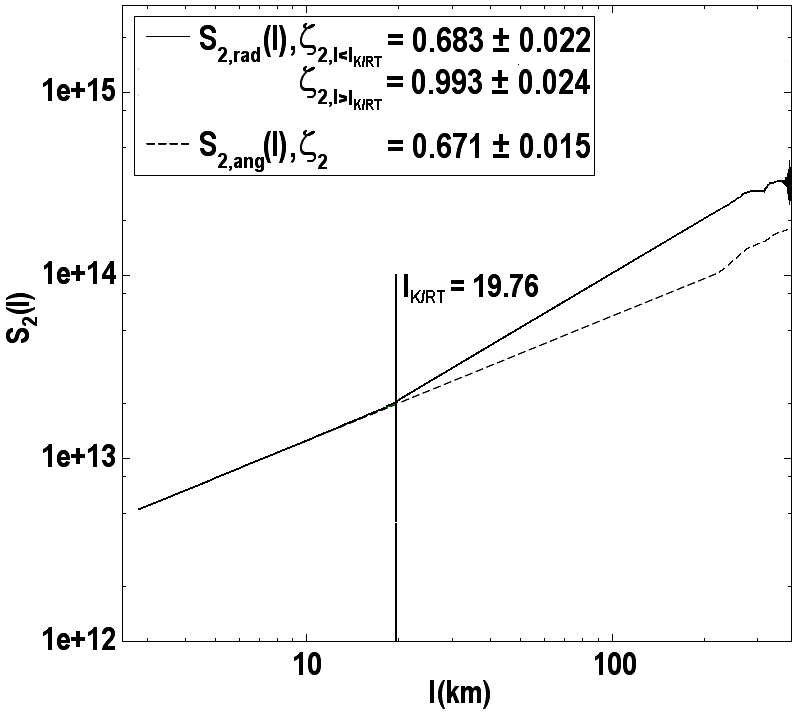}}
        \subfigure[t = 0.5 seconds]
        {\includegraphics[width=5.20cm]{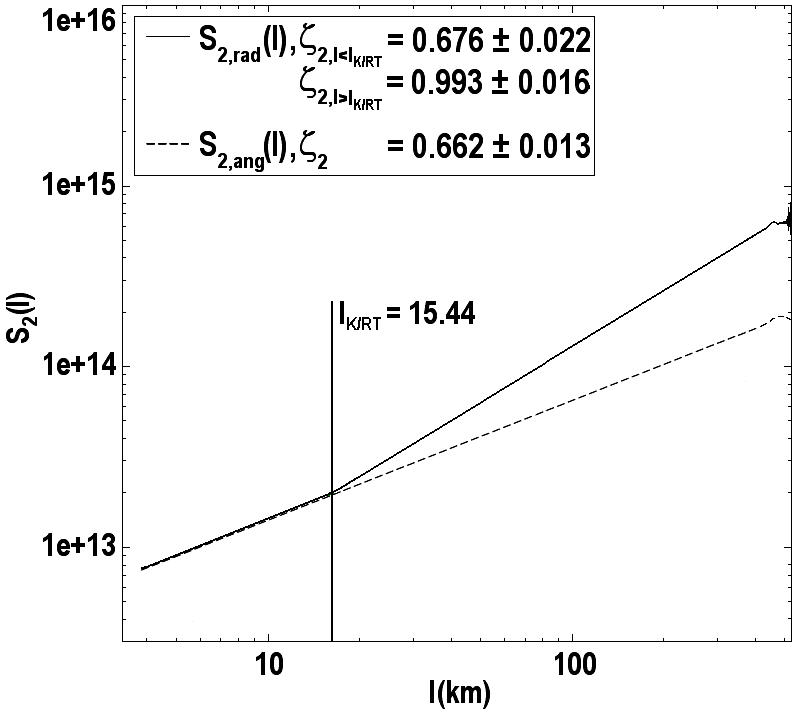}}
        \subfigure[t = 0.6 seconds]
        {\includegraphics[width=5.20cm]{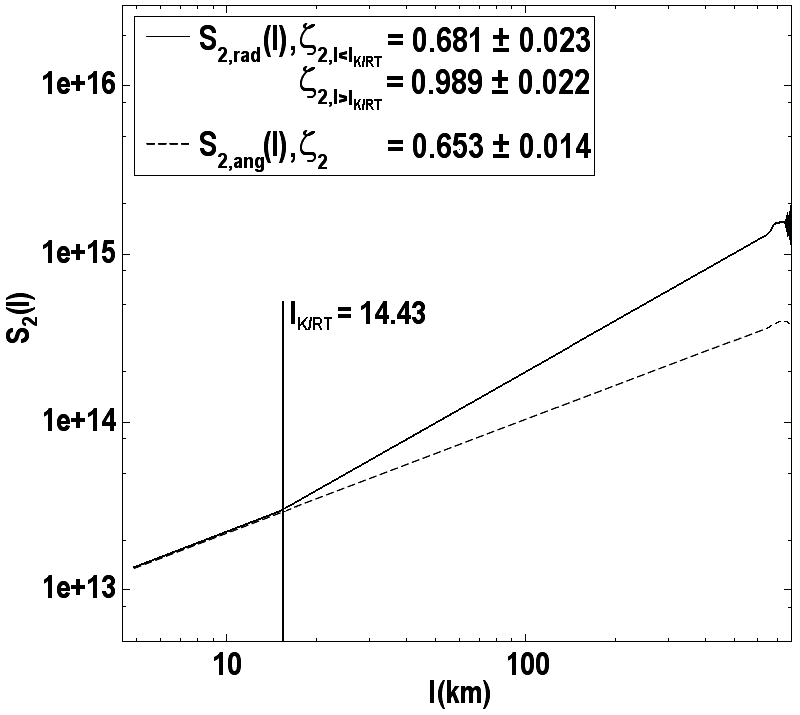}}
        \subfigure[t = 0.7 seconds]
        {\includegraphics[width=5.20cm]{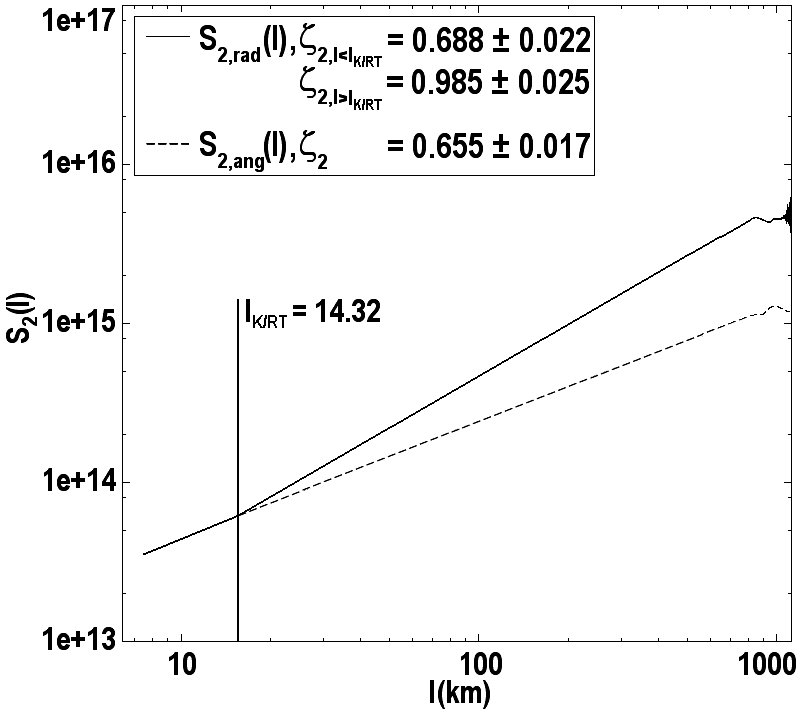}}
        \subfigure[t = 1.0 seconds]
        {\includegraphics[width=5.20cm]{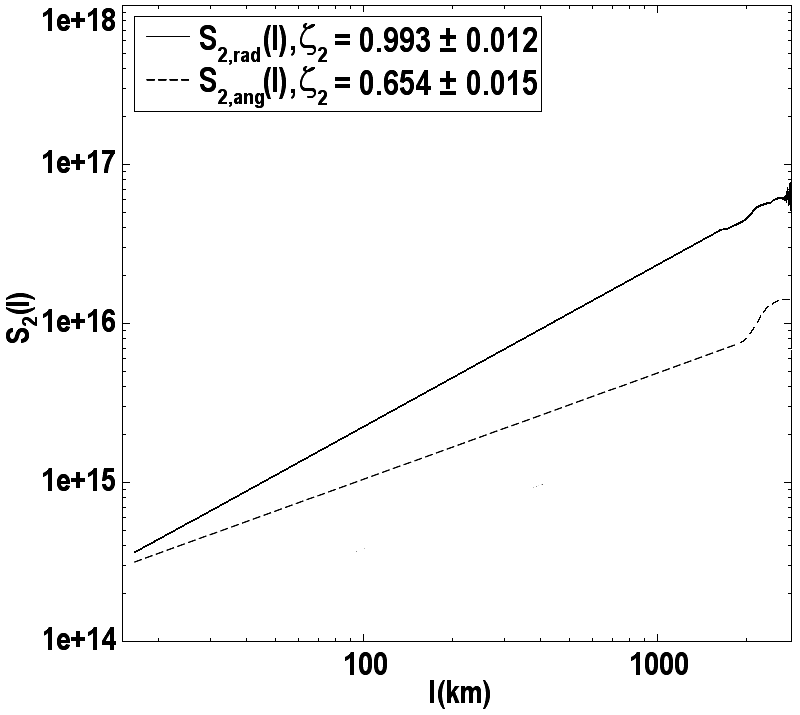}}
\caption{Radial (solid curve) and angular (dashed curve) 2nd-order structure function with the determined transition length scale.} 
\label{Fig:2}
\end{figure*}

\begin{deluxetable*}{cccccccc}
\label{Tab:1}
\tablecaption{Relative scaling exponents $Z_{p}$ at three different instants of time}
\tablewidth{\textwidth}
\tablehead{
& \colhead{$t [s]$} & \colhead{$p=1$} & \colhead{$p=2$} & \colhead{$p=3$} &
\colhead{$p=4$} & \colhead{$p=5$} & \colhead{$p=6$}
}
\startdata
& $0.5$ & $0.344\pm 0.016$ & $0.676\pm 0.027$ & $1.0\pm 0.037$ & $1.309\pm 0.050$ & $1.597\pm 0.058$ & $1.860\pm 0.068$\\
$Z_{p,\mathrm{rad}}\ \mbox{for}\ \ell<\ell_{\mathrm{K/RT}}$ & $0.6$ & $0.341\pm 0.016$ & $0.676\pm 0.030$ & $1.0\pm 0.046$ & $1.304\pm 0.057$ & $1.593\pm 0.071$ & $1.850\pm 0.082$\\
& $0.7$ & $0.347\pm 0.019$ & $0.676\pm 0.029$ & $1.0\pm 0.038$ & $1.302\pm 0.052$ & $1.584\pm 0.060$ & $1.842\pm 0.074$\\ \tableline 
& $0.5$ & $0.339\pm 0.009$ & $0.674\pm 0.015$ & $1.0\pm 0.021$ & $1.315\pm 0.027$ & $1.615\pm 0.033$ & $1.900\pm 0.038$\\
$Z_{p,\mathrm{rad}}\ \mbox{for}\ \ell>\ell_{\mathrm{K/RT}}$ & $0.6$ & $0.338\pm 0.011$ & $0.674\pm 0.019$ & $1.0\pm 0.027$ & $1.316\pm 0.033$ & $1.613\pm 0.042$ & $1.896\pm 0.050$\\  
& $0.7$ & $0.338\pm 0.011$ & $0.674\pm 0.019$ & $1.0\pm 0.027$ & $1.315\pm 0.035$ & $1.615\pm 0.043$ & $1.898\pm 0.050$\\ \tableline
& $0.5$ & $0.336\pm 0.008$ & $0.670\pm 0.016$ & $1.0\pm 0.024$ & $1.323\pm 0.034$ & $1.636\pm 0.042$ & $1.920\pm 0.050$\\  
$Z_{p,\mathrm{ang}}$ & $0.6$ & $0.335\pm 0.013$ & $0.668\pm 0.018$ & $1.0\pm 0.026$ & $1.319\pm 0.034$ & $1.625\pm 0.045$ & $1.919\pm 0.050$\\  
& $0.7$ & $0.335\pm 0.009$ & $0.670\pm 0.016$ & $1.0\pm 0.022$ & $1.323\pm 0.030$ & $1.635\pm 0.035$ & $1.934\pm 0.041$
\enddata
\label{powerexponents}
\end{deluxetable*}

\section{Results}
\label{Results}
\subsection{Energy spectra}
\label{ResultsI}

We plot the energy spectra as functions of the normalized wave number
$k_{\mathrm{n}} = 512\Delta_0(t)k/\pi$ for $16\le k_{\mathrm{n}}\le
512$, where $\Delta_0(t)$ is the size of the cells in the uniform part
of the grid at time $t$. We have $\Delta_0(t)=2.93$ and
$14.69\,\mathrm{km}$ at times $t=0.5$ and $1.0$ seconds,
respectively. Note that wave numbers $k_{\mathrm{n}}<32$ are obscured
by data windowing (see Section~\ref{Analysis}). The computed energy
spectrum functions at 0.5 seconds are shown as black curves in
Fig. \ref{Fig:1} (a). The thin gray lines corresponds to power-law
fits and the dashed line indicates the expected spectrum according to the
Kolmogorov theory with an exponent $\beta = 5/3$. For the longitudinal
spectrum function, we find $\beta \approx 1.64$ for high wave numbers,
which is in good agreement with the Kolmogorov theory. The longitudinal spectrum
possibly stiffens toward lower wave numbers, but the accuracy of the
computed spectra does not allow for a conclusive result. We exclude
the lower part of wavenumbers from the fit, because the slope becomes steeper than a Kolmogorov spectrum for $k_\mathrm{n}\lesssim 64$. The transversal
spectrum is slightly steeper ($\beta \approx 1.71$), but close to
a Kolmogorov spectrum for the whole range of wave numbers. 
These results were corroborated by the computation of 
second-order structure functions with much higher accuracy (see section~\ref{ResultsII}). 
We also note that the Kolmogorov scaling for higher wave numbers agrees with the findings of
\citet{RoepHille07}, where turbulence energy spectra were computed
without splitting the velocity field. This is in accordance with the expectation that
the radial expansion will mostly affect low wave numbers modes. At $t=1.0$
seconds, on the other hand, the longitudinal spectrum has an exponent
$\beta \approx 1.97$ over the entire range of wave numbers. In
contrast, the transverse spectrum is much shallower. The exponent
$\beta= 2$ corresponds to RT scaling, because $E(k)\propto k^{-2}$
implies $\delta v(\ell)\propto \ell^{1/2}$. Consequently, it appears
that the velocity component in the direction of gravity is dominated
by the RT instability even at the smallest numerically resolved scales
in the late phase of the explosion, while Kolmogorov scaling is found
for the velocity component perpendicular to gravity. 

\subsection{Velocity structure functions}
\label{ResultsII}

\begin{figure}[t!]
\includegraphics[width=0.45\textwidth]{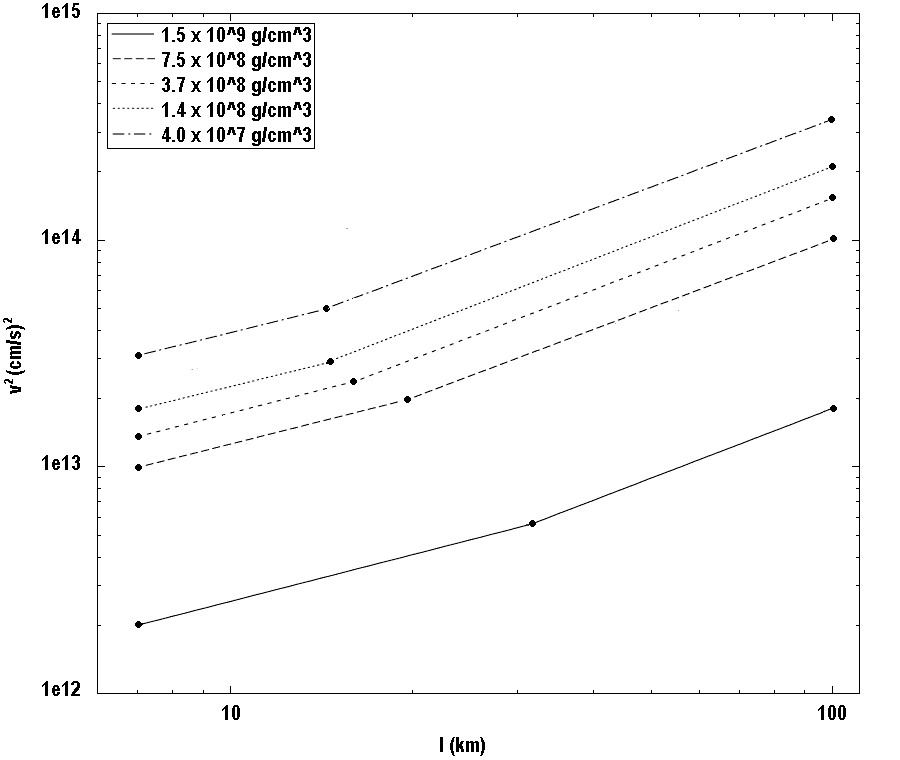}
\caption{2nd order radial structure functions with the transition
  length (big dots) and corresponding density on the flame front.} 
\label{Fig:3}
\end{figure}

For the computation of the structure functions following equation
(\ref{Eq:5a}) and (\ref{Eq:5b}), we chose a spherical region in which
90\% of the material was burned. This choice of the region resulted
from the requirement of encompassing the bulk of turbulence at a given
time, while excluding the outer, non-turbulent regions of the white
dwarf. Fig. ~\ref{Fig:2} shows double-logarithmic plots of the
radial (solid curve) and angular (dashed curve) structure functions of
second order at different times. In all cases up to $t=0.7$ seconds, a
scaling exponent $\zeta_{2}$ close to $2/3$ is found for the range of
length scales $\ell\lesssim 10\,\mathrm{km}$. Moreover,
$S_{p,\mathrm{rad}}(\ell)\approx S_{p,\mathrm{ang}}(\ell)$, which
indicates isotropy at small length scales. For larger length scales,
on the other hand, the radial structure functions
$S_{p,\mathrm{rad}}(\ell)$ obey a scaling law with an exponent $\zeta_{2,\mathrm{rad}}\approx 1$,
whereas $S_{p,\mathrm{ang}}(\ell)$ still follows Kolmogorov
scaling. From $\zeta_{2,\mathrm{rad}}\approx 1$, it follows that $v'(\ell)\propto\ell^{1/2}$.
As outlined in section~\ref{Analysis}, this corresponds to RT scaling. For this reason, the change of slope of the radial structure function
indicates a transition from the inertial-range turbulence cascade to the regime of RT instabilities
 at a length scale
$\ell_{\mathrm{K/RT}}$ approaching $\approx 14\,\mathrm{km}$ in the
course of the explosion. Thus, our analysis confirms the estimate by
\citet{NieWoos97}. As a result of the overall expansion of the
co-moving grid, the transition length drops below the numerical
resolution of the simulation after $\sim$ $0.7$ to $1.0$
seconds. The time evolution of $\ell_{\mathrm{K/RT}}$ is further
illustrated in Fig. \ref{Fig:3}, where the second order structure
functions are plotted for all data sets, for which $\ell_{\mathrm{K/RT}}$ is
numerically resolved.
While in Fig.~\ref{Fig:2} the range of length scales is adjusted to the size
of the co-moving grid, a fixed range of length scales is used in Fig.~\ref{Fig:3}.
Additionally, the typical mass density in the vicinity of the flame front is specified
for each instant of time. In agreement with Fig. 1 of \citet{NieWoos97}, Fig. 2 and 3 show that $\ell_{\mathrm{K/RT}}$ becomes smaller with decreasing density (and advancing time). Remarkably, it appears that $\ell_{\mathrm{K/RT}}$ approaches an asymptotic value,
but we cannot investigate this behavior for $t > 0.8$ seconds. To obtain more properties of $\ell_{\mathrm{K/RT}}$ additional high-resolved numerical simulations are needed,
in which $\ell_{\mathrm{K/RT}}$ can be tracked for a longer time.

\begin{figure*}[t!]
\centering
        \subfigure[t = 0.5 seconds, radial]
        {\includegraphics[width=5.20cm]{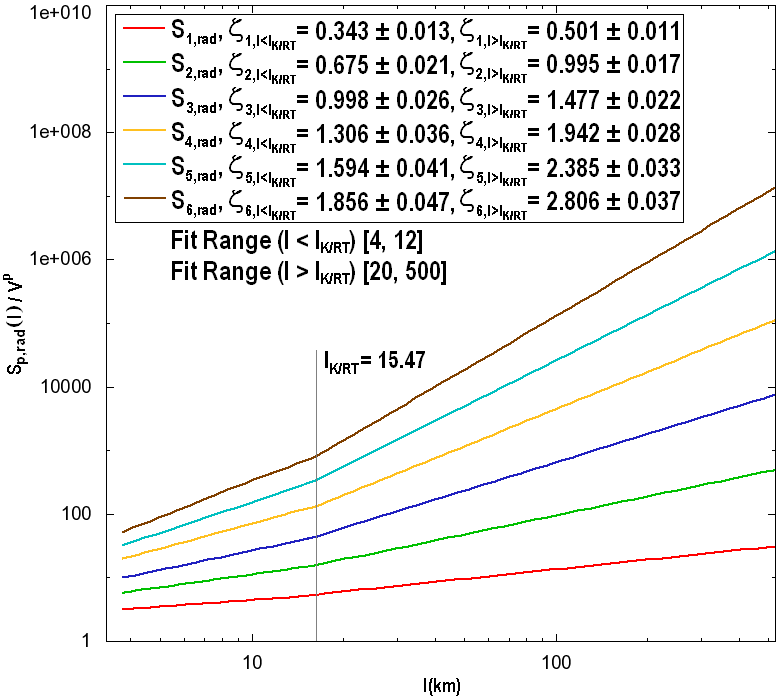}}
        \subfigure[t = 0.6 seconds, radial]
        {\includegraphics[width=5.20cm]{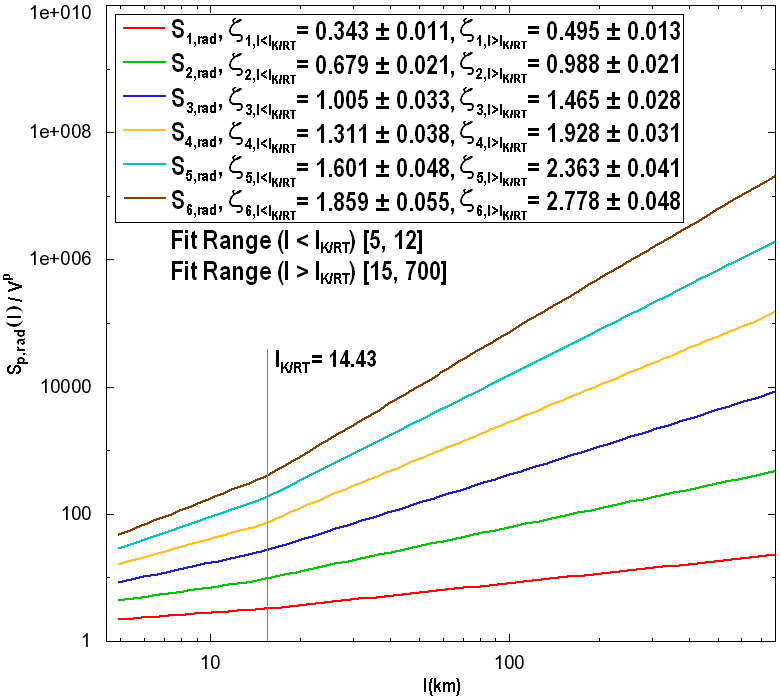}}
        \subfigure[t = 0.7 seconds, radial]
        {\includegraphics[width=5.20cm]{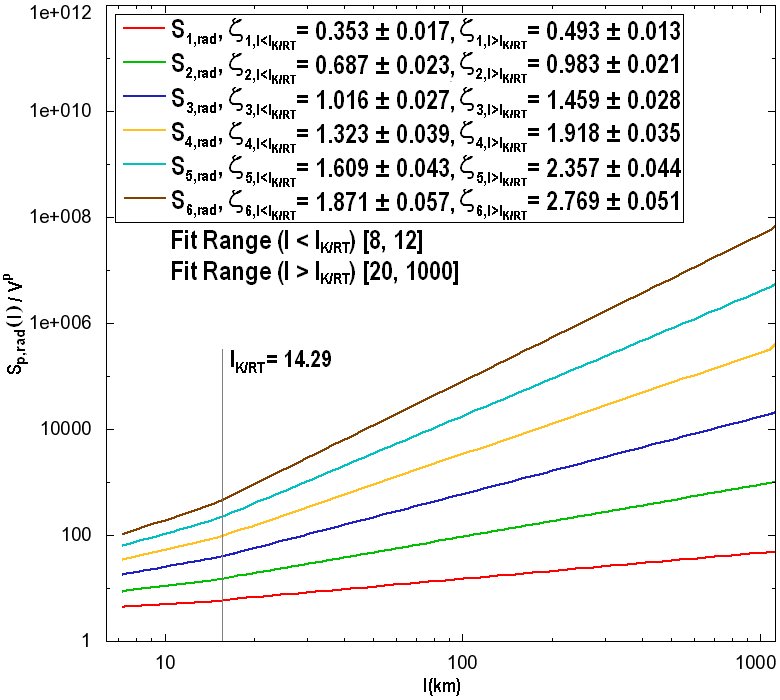}}
        \subfigure[t = 0.5 seconds, angular]
        {\includegraphics[width=5.20cm]{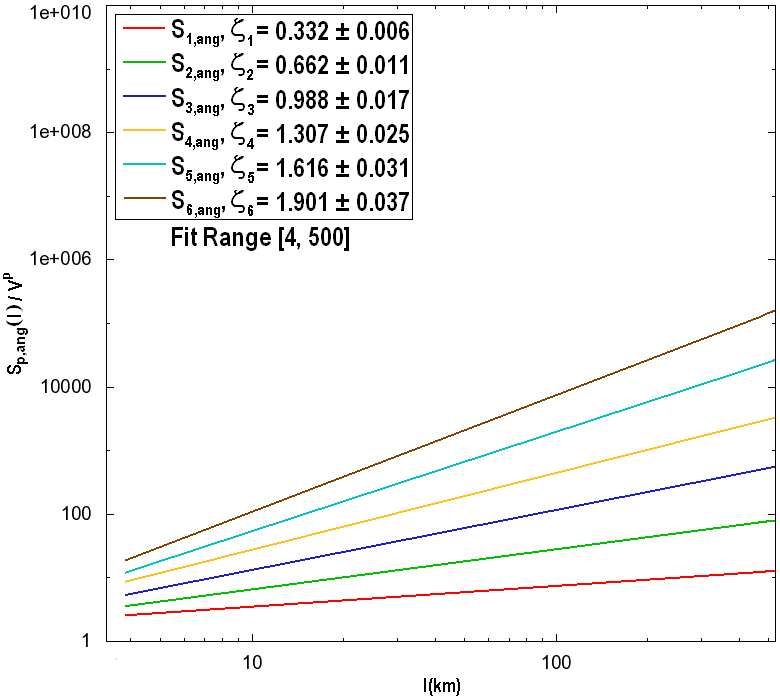}}
        \subfigure[t = 0.6 seconds, angular]
        {\includegraphics[width=5.20cm]{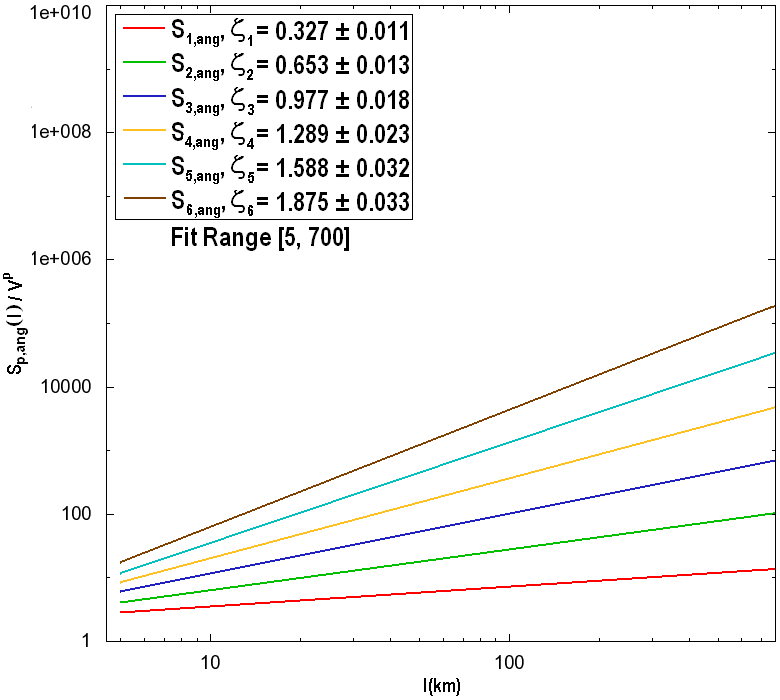}}
        \subfigure[t = 0.7 seconds, angular]
        {\includegraphics[width=5.20cm]{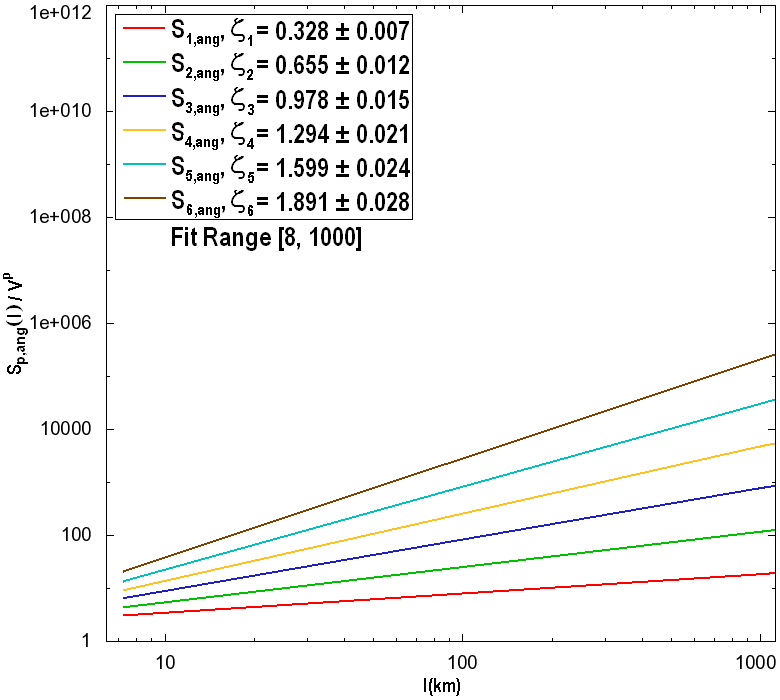}}
\caption{Radial and angular structure function up to the sixth order with the corresponding scaling exponents. For the radial structure functions, the transition length from Kolmogorov to Rayleigh-Taylor scaling $\ell_{\mathrm{K/RT}}$ is indicated.} 
\label{Fig:4}
\end{figure*}

In Figure~\ref{Fig:4}, the structure functions up to the sixth order
are plotted for $t=0.5$, $0.6$ and $0.7$ seconds. As one can see from
the scaling exponents listed in Figure~\ref{Fig:3}, $\zeta_{3}$ 
is close to unity for the angular structure functions and for the
radial structure functions in the subrange $\ell<\ell_{\mathrm{K/RT}}$. 
This result is consistent with the Kolmogorov  theory \citep[see][]{Frisch}.
For $\ell>\ell_{\mathrm{K/RT}}$, on the other hand, 
$\zeta_{3,\mathrm{rad}}\approx 1.5$. Remarkably, it appears that
for all $p\le 6$ the slopes of the radial structure functions are
steeper by a factor of $1.5$ at length 
scales greater than $\ell_{\mathrm{K/RT}}$. This suggests that the
ratio $Z_{p}:=\zeta_{p}/\zeta_{3}$ is approximately equal for
$S_{p,\mathrm{rad}}(\ell<\ell_{\mathrm{K/RT}}) $ and
$S_{p,\mathrm{rad}}(\ell>\ell_{\mathrm{K/RT}}) $. Indeed, the
\emph{relative} scaling exponents 
of the radial structure functions in both subranges are quite close
(see Table~\ref{powerexponents}). 
There are theoretical arguments \citep{Dubr94} as well as numerical
investigations \citep{Benzi93} in support of the fundamental
significance of relative scaling exponents. But it has not been noticed
before that RT-driven velocity fluctuations exhibit statistical
properties that are equivalent to the properties of isotropic,
inertial-range turbulence in terms of relative scalings. It is known
that the calculation of higher-order exponents becomes increasingly
uncertain due to sampling errors \citep{Frisch}. For this reason, we do not consider the
relatively high discrepancies between the results for $p\ge 5$
to be significant. Comparing the relative
scaling exponents of the radial and the angular structure functions in
the subrange $\ell>\ell_{\mathrm{K/RT}}$, on the other hand, we find
very good agreement. We will concentrate on these scaling exponents in
the following. 

\begin{figure}[ht]
\centering
\includegraphics[width=0.42\textwidth]{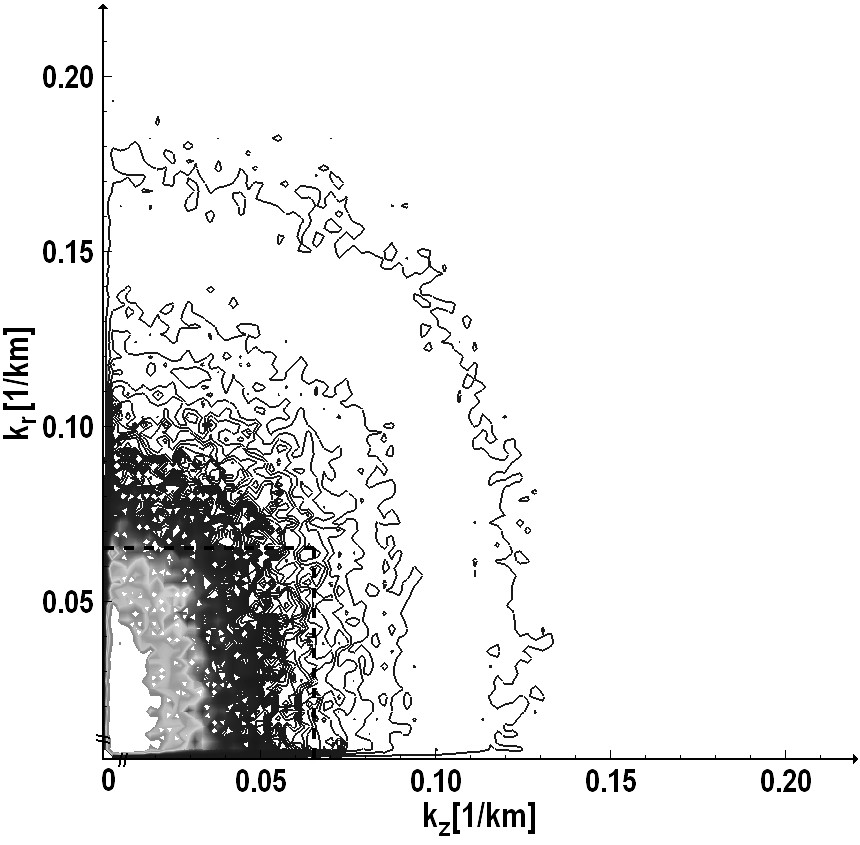}
\caption{Contour plots of the Fourier-transformed velocity differences
  inside a small box 
intersected by the flame front at 0.5 seconds.}
\label{Fig:5}
\end{figure}

\subsection{Local isotropy of the velocity fluctuations}
\label{ResultsIII}

The contour lines of the Fourier-transformed velocity field at 0.5
seconds inside a box as described in Section~\ref{Analysis} are
plotted in Fig~\ref{Fig:5}. The box is positioned such that $50\%$
of the enclosed matter is burned. The thick dashed line indicate the
wave number $2\pi/\ell_{\mathrm{K/RT}}$ corresponding to the
transition length obtained from the calculation of the radial
structure functions. For smaller wave numbers (i.~e.,
$\ell\gtrsim\ell_{\mathrm{RT}})$, the contours of
$\delta\hat{\mathbf{v}}(\mathbf{k})$ are clearly anisotropic. One can see
that a given velocity difference in the $z$-direction spans a smaller
range of wave numbers than in the directions perpendicular to the
$z$-axis. This corresponds to a steeper slope of the velocity
fluctuations in the radial direction, which is approximately given by
the $z$-direction. Toward higher wave numbers 
($\ell\lesssim\ell_{\mathrm{RT}})$, the anisotropy of the contours
decreases, but we do not find perfect isotropy. This might indicate
residual small-scale anisotropy in the vicinity of the flame front,
which can be caused by the intermittency of turbulence. However, it
could also be a spurious effect due to the misalignment between the
$z$-axis and the radial direction at off-center positions within the
box.

\section{Conclusion}

We investigated the scaling properties of turbulence in a
high-resolution simulation of a Type Ia supernova based on the pure
deflagration model \citep{RoepHille07}. Both energy spectrum functions
and structure functions of second order were computed. The results of
this study are as follows: 

\begin{enumerate}
\item The velocity fluctuations in the radial direction, i.~e., the
  direction of gravity, follow Kolmogorov scaling at length scales
  smaller than a certain transition length $\ell_{\mathrm{K/RT}} \sim
  10\,\mathrm{km}$. Only at length scales greater than
  $\ell_{\mathrm{K/RT}}$, the radial velocity fluctuations are
  dominated by the scaling law of the Rayleigh-Taylor
  instability. This behaviour was predicted by \citet{NieWoos97}. 
\item The velocity fluctuations in angular directions, i.~e.,
  perpendicular to gravity, obey Kolmogorov scaling over the entire
  range of numerically resolved length scales. 
\item For $\ell\lesssim\ell_{\mathrm{K/RT}}$, the magnitudes of the
  radial and angular velocity fluctuations are nearly equal. For this
  reason, small-scale turbulence appears to be statistically
  isotropic. 
\item Fourier analysis of the velocity fluctuations in a small region
  near the flame surface allows for slight residual anisotropies at
  the smallest resolved scales. 
\end{enumerate}

As regards the interpretation of our results, a possible cause for
concern is that Kolmogorov scaling in the radial direction is only
found for a relatively narrow range of length scales greater than the
numerical cutoff length. It is known that these scales are affected by
numerical dissipation and, particularly, by the bottleneck effect
\citep{SchmHille06}. Thus, the flattening of the turbulence energy
spectrum might be artificial. However, in this case, no
significant flattening should be observed for the corresponding
structure functions which are much less affected by the bottleneck
effect. The scaling laws implied by the 
radial energy spectra and structure functions are fully consistent at
small length scales and the transversal spectra show no flattening at
all. In consequence, we are confident that the scaling laws are
genuine. A possible explanation for the absence of the bottleneck
effect is that turbulence does not reach a statistically stationary
state in a supernova explosion. 

Yet another issue is that Kolmogorov scaling might be enforced by the
subgrid scale model that was used in the simulation. However, other
than the RT-based models used by \citet{GamKhok03}, this SGS model
does neither presume any given scaling of turbulence nor does it
imprint such a scaling on the numerically resolved flow. A potential
problem for the SGS model is the possible lack of isotropy near the
flame surface. However, there is certainly no pronounced anisotropy at
the smallest resolved scales if $\Delta_{0}(t)<\ell_{\mathrm{K/RT}}$,
and statistical isotropy is found for the bulk of turbulent
regions. Only in the late phase of the explosion, when the transition
length $\ell_{\mathrm{K/RT}}$ becomes smaller than the numerical
resolution and the resolved small-scale turbulence definitely becomes
anisotropic, the notion of SGS turbulence energy cannot be strictly
justified. One should note, however, that this point more or less
coincides with the time when a deflagration-to-detonation transition
is expected to occur \citep{GamKhok05,RoepNie07}. Apart from that, the
production of turbulence energy by unresolved buoyancy effects is
heuristically included in the SGS model. In conclusion, the SGS
turbulence energy model by \cite{SchmNie06} applies to the major part
of the explosive burning in the deflagration phase of a Type Ia
supernova explosion, but there is no regime for which a pure
RT-scaling model holds. 

After settling the issue of turbulence scaling in the deflagration
phase of a Type Ia supernova in the present article, we mention that
the occurrence of deflagration-to-detonation transitions can be constrained on the
basis of the deflagration model. For the DDT mechanism to operate,
strong turbulence is necessary in late phases of the burning \citep[e.g.][]{woosley2007a}. 
\citet{roepke2007d} found that this may indeed be realized in
deflagration models of SNe~Ia with low (but not vanishing)
probability. In order to better quantify this intermittency effect, higher-order structure
functions have to be computed. Fitting intermittency models to the
numerically determined scaling exponents \citep[as proposed
by][]{Pan08}, the probability of strong
turbulent velocity fluctuations at any
instant of time can be estimated. This analysis will be presented in a future
publication.

\acknowledgments{The research of F.K.R.\ is supported
  through the Emmy Noether Program of the German Research Foundation (DFG;
  RO~3676/1-1).}


\end{document}